# An Ontological Interpretation of Photon Wave-Particle Duality via Complex-Space Trajectories

**Shiang-Yi Han[1] and Ciann-Dong Yang[2,*]**

[1] Department of Aerospace Engineering, Tamkang University, New Taipei City 251, Taiwan, R.O.C.

[2] Department of Astronautics and Aeronautics, National Cheng Kung University, Tainan, 701, Taiwan, R.O.C.

E-mail: 168936@mail.tku.edu.tw

*cdyang@ncku.edu.tw

* Corresponding author

## Abstract

Wave–particle duality remains a central interpretational challenge in quantum theory. In this work, we develop a trajectory-based description of photon dynamics formulated in an extended complex space within the relativistic quantum Hamilton–Jacobi framework. In this approach, photon motion is represented by complex trajectories whose real projections describe propagation, while imaginary components encode oscillatory structure. We show that momentum eigenstates correspond to straight-line trajectories with uniform propagation at the speed of light, whereas superposition states lead to nontrivial quantum potentials and oscillatory motion in the complex plane. Extending the analysis to complex two-dimensional space reveals richer dynamical behavior, including propagating wave-like trajectories and standing-wave-like patterns in real projections. Energy–momentum consistency is verified through an internal coherence analysis based on projected standing-wave wavelengths. Rather than introducing new dynamical laws or additional physical dimensions, the complex space is employed as an interpretational framework that renders wave-like and particle-like aspects as complementary projections of a single underlying motion. The results suggest a unified geometric perspective on wave–particle duality, while remaining fully compatible with standard quantum mechanics.

## 1. Introduction

Wave-particle duality has accompanied the development of quantum theory since its inception. From Planck's quantization of radiation and Einstein's photono hypothesis to de Broglie's matter wave, the coexistence of particle-like and wave-like descriptions has played a central role in shaping modern physics. Despite this historical success, wave-particle duality remians primarily an interpretational issue rather than a formal one. Standanrd quantum mechanics accurately predicts experimental outcomes through the probabilistic structure of the wave function, yet deliberately refrains from providing a detailed account of the underlying physical reality between measurements. This instrumentalist stance, while operationally effective, leaves open fundamental questions concering the ontology of quantum systems. In particular, it remains unclear whether wave-particle duality reflects a fundamental dualism of physical entities or merely different representaional aspects of a single underlying physical process.

Over the past decades, several realist and trajectory-based approaches have sought to address this issue while preserving empirical equivalence with standard quantum mechanics. Examples include Bohmian mechanics, quantum Hamilton-Jocobi formulation, and weak measurement reconstructions of average quantum trajectories [1-7]. These approaches share a common motivation: to provide an explicit ontological picture of microscopic dynamics without sacrificing predictive success.

In recent years, independent experimental developments have provided further motivation for considering complex-valued structures in the description of quantum dynamics [8-12]. In particular, weak-measurement experiments have demonstrated that physically accessible quantities—such as weak values of position, momentum, and velocity—can take intrinsically complex values. These complex weak values are not formal artifacts: their real and imaginary components have been separately measured and shown to govern distinct physical effects, including average trajectories, probability flow, and measurement backaction [13-16]. As a result, quantum dynamics can be operationally reconstructed in terms of complex-valued trajectories inferred from experimental data.

From this perspective, the extension of configuration space into the complex domain is not introduced as a speculative modification of physical spacetime, but rather as a natural interpretational response to experimentally observed complex quantities [17-21]. Weak-measurement reconstructions suggest that quantum evolution effectively explores a complexified space, whose imaginary components encode information inaccessible to strong measurements but nevertheless influence observable outcomes . Our former works adopts this viewpoint and treats complex space as an ontological bookkeeping device that renders such experimentally supported complex dynamics geometrically explicit [22-27]. In this sense, the complex-trajectory framework employed here may be viewed as a theoretical completion of structures already implicit in weak-measurement phenomenology.

In this article, we explore how wave-particle duality can be coherently interpreted when particle motion is represented within an extended, complex space on the based of relativistic quantum Hamiton-Jacobi formulation [28]. In this framework, particle-like and wave-like behavior emerge as different projections of a single underlying complex trajectory, offering a unified ontological perspecctive on photon dynamics. The interpretation adopted in this article is explicitly ontological in nature. Its purpose is to describe what is assumed to exist at the microscopic level rather than merely providing operational rules for predicting measurement outcomes. As the same time, this ontological commitment is deliberately conservative: the formal structure of quanatun mechanics is left entirely intact, and no additional hidden variables are introduced beyond those implicit in the wave function. The extension of space into complex domain in not postulated as a new physical spacetime. Instead, it is employed as an interpretational framework within which quantum phenomena may be coherently visualized.

Within this perspective, the wave function encodes information about the geometry of particle motion in an extended complex space, whose complex structure gives rise to observable wave phenomena when projected onto real coordinates. This viewpoint bears conceptual similarities to Bohmian mechanics in its emphasis on trajectoreis and realism. However, it differs in a crucial respect: wave behavior is attrubuted not to an indepdent guiding equation acting in real space, but to the intrinsic geometry of the complexified Hamilton-Jacobi formulation itself. The associated quantum potential arises as a structural feature rather than as an addition al physical interaction.

Within the complex-trajectory framework, superposition-induced quantum potentials can give rise not only to propagating wave-like motion but also to bounded dynamics in the extended complex space. In particular, when the photon occupies a fully bounded mode in multiple directions, its motion becomes confined within an effective potential well in complex space. Although the full dynamics unfold in a higher-dimensional complex domain, their physical manifestations can be analyzed through projections onto real coordinate planes, where standing-wave-like patterns naturally emerge. This provides a geometric interpretation of standing waves as projected images of bounded complex trajectories rather than as independent physical entities. Importantly, such bounded and standing-wave-like behavior does not violate fundamental conservation laws. Energy–momentum consistency is maintained within the relativistic quantum Hamilton–Jacobi formulation, even when the projected motion appears purely wave-like. In this sense, the familiar relations associated with particle and wave descriptions—such as Einstein's photon energy and the de Broglie wavelength—are not treated as separate postulates but arise as mutually consistent constraints within a single dynamical picture. The complex-trajectory framework therefore offers a coherent geometric setting in which particle-like transport and wave-like structure are unified as different projections of the same underlying motion.

This paper is organized as follows. Section 2 reviews the relativistic quantum Hamilton–Jacobi formulation and demonstrates how the Klein–Gordon equation can be derived within this framework without introducing operator postulates. In Section 3, we analyze the motion of a single free photon in the complex plane and show how the associated complex quantum potential gives rise to wave-like trajectories. Section 4 extends the analysis to photon dynamics in complex $x$–$y$ space, where richer propagating and standing-wave behaviors emerge. In this context, the connection between Einstein's photon energy and the de Broglie relation is clarified through the geometry of complex trajectories. Section 5 concludes with a summary of the results and a discussion of their implications.

## 2. Relativistic Quantum Hamilton-Jacobi Framework

To formulate a trajectory-based ontological interpretation consistent with quantum mechanics, we adopt a relativistic quantum Hamilton-Jacobi framework in which both the space coordinates and the action function are complex values. We introduce the complex action function $S(t, q^\alpha)$ through its relation to the wave function

$$S(t, q^\alpha) = -i\hbar \ln \Psi(t, q^\alpha), \qquad q^\alpha \in \mathbb{C} \tag{1}$$

where $q^\alpha = (ct, x, y, z)$ denotes complex spacetime coordinates. The relativistic quantum Hamilton-Jacobi equation is writen as [28]:

$$\frac{\partial S}{\partial \tau} + \frac{c^2}{E_0}\left[\left(\frac{\partial S}{\partial q^0}\right)^2 - \sum_{j=1}^{3}\left(\frac{\partial S}{\partial q^j}\right)^2\right] + \frac{c^2\hbar}{E_0} i \left(\frac{\partial^2 S}{\partial q^0 \partial q^0} - \sum_{j=1}^{3}\frac{\partial^2 S}{\partial q^j \partial q^j}\right) = 0, \tag{2}$$

where $\tau$ is the proper time and $E_0 = m_0 c^2$. Substitution of Eq. (1) into Eq. (2) yields the Klein-Gordon equation

$$\left(\frac{1}{c^2}\frac{\partial^2}{\partial t^2} - \nabla^2 \frac{m_0^2 c^2}{\hbar^2}\right)\Psi(t, x, y, z) = 0, \tag{3}$$

demonstrating formal equivalence with standard relativistic quantum mechanics.

The canonical moementum is defined by

$$p_\alpha = \frac{\partial S}{\partial q^\alpha}, \tag{4}$$

and the equations of motion follow from

$$\frac{dq^\alpha}{d\tau} = \frac{\partial H}{\partial p_\alpha}, \qquad \frac{dp_\alpha}{d\tau} = -\frac{\partial H}{\partial q^\alpha}. \tag{5}$$

These equations define complex trajectories that serve as ontological representations of photon motion. In this formulation, relativistic wave dynamics follow directly from classical action principles expressed in complex spacetime. By substituting the separable solution

$$\Psi(t,x,y,z) = T(t)X(x)Y(y)Z(z), \tag{6}$$

into Eq. (3), we have the relativistic energy–momentum relation

$$k_0^2 = m_0^2c^4 + c^2\hbar^2(k_1^2+k_2^2+k_3^2) = E_0^2 + c^2(p_1^2+p_2^2+p_3^2), \tag{7}$$

together with four decoupled equations,

$$\frac{1}{T}\frac{d^2T}{dt^2} = -\frac{k_0^2}{\hbar^2}, \qquad \frac{1}{X}\frac{d^2X}{dx^2} = -k_1^2, \qquad \frac{1}{Y}\frac{d^2Y}{dy^2} = -k_2^2, \qquad \frac{1}{Z}\frac{d^2Z}{dz^2} = -k_3^2. \tag{8}$$

Here $k_0$ represents the total energy, while $k_j$ ($j = 1,2,3$) are related to the spatial momentum components via $p_j = \hbar k_j$. Eq. (8) therefore expresses energy conservation for a relativistic particle within the present complex-spacetime formulation. The equations of motion of a free relativistic particle follow from Eq. (4)

$$\frac{dq^0}{d\tau} = \frac{c^2}{E_0}p_0, \qquad \frac{dq^j}{d\tau} = -\frac{c^2}{E_0}p_j, \qquad j = 1,2,3, \tag{9}$$

which, in the local frame, yield

$$\frac{\partial x^j}{\partial t} = -\frac{cp_j}{p_0} = -\frac{c^2\hbar k_j}{k_0}, \qquad j = 1,2,3. \tag{10}$$

The corresponding particle velocity is

$$v = \sqrt{\frac{c^2\hbar^2(k_1^2+k_2^2+k_3^2)}{k_0^2}}, \tag{11}$$

which is always less than the speed of light $c$. In the massless limit $m_0 \to 0$, one finds

$$k_0^2 = c^2\hbar^2(k_1^2+k_2^2+k_3^2), \tag{12}$$

providing the basis for describing photon motion and its connection to wave propagation within this framework.

## 3. Single Photon Motions in Complex Plane

We consider a single free photon and analyze its motion in the complex coordinate $x = x_R + ix_I \in \mathbb{C}$ by solving the equation of motion derived from Eq. (4). In the local frame, the equation of motion can be written as

$$\frac{dx}{dt} = -\frac{cp_1}{p_0} = -\frac{ic\hbar}{k_0}\frac{\partial}{\partial x}\ln\,\Psi(t,x), \qquad x \in \mathbb{C}, \tag{13}$$

where the momentum–wavefunction relation $p_1 = -i\hbar\,\partial_x \ln\Psi$ has been used. The photon wave function is taken as a superposition of right- and left-propagating plane-wave components,

$$\Psi(t,x) = e^{i(k_0/\hbar)t}(C^+e^{ik_1x} + C^-e^{-ik_1x}), \tag{14}$$

where $k_0$ characterizes the total energy, $k_1$ is the spatial wave number ($p_1 = \hbar k_1$), and $C^+$, $C^-$ are the amplitudes of the $+x\ (right)$ and $-x\ (left)$ eigenstates, respectively. Throughout this article we consider forward time evolution, $dt/d\tau > 0$. Introducing the dimensionless variables

$$\bar{t} = \frac{k_0}{\hbar}t, \qquad \bar{x} = k_1 x, \qquad A = \frac{C^+}{C^-},$$

Eq. (13) can be rewritten as

$$\frac{d\bar{x}}{d\bar{t}} = \frac{Ae^{i\bar{x}} - e^{-i\bar{x}}}{Ae^{i\bar{x}} + e^{-i\bar{x}}} = \dot{\bar{x}}_R + i\dot{\bar{x}}_I \in \mathbb{C}. \tag{15}$$

For an eigenstate, $A = 0$ or $A \to \infty$, $|\, d\bar{x}/d\bar{t}\,| = 1$, and therefore $|\, dx/dt\,| = c$. The photon trajectory is a straight line, consistent with particle-like propagation and the special relativity. For a superposition state $0 < |\,A\,| < \infty$, the velocity becomes complex and the trajectory $\bar{x}(\bar{t}) = \bar{x}_R(\bar{t}) + i\bar{x}_I(\bar{t})$ is generally non-straight, exhibiting oscillatory structure in the complex plane. Integrating Eq. (15) yields the implicit solution

$$Ae^{i\bar{x}} - e^{-i\bar{x}} = De^{-i\bar{t}}, \tag{16}$$

where the integration constant $D$ is determined by the initial conditions $\bar{x}(0)$ and $\dot{\bar{x}}(0)$. A convenient expression is

$$D = \frac{2\dot{\bar{x}}(0)}{1 - \dot{\bar{x}}(0)}\, e^{-i\bar{x}(0)}. \tag{17}$$

From Eqs. (16) and (17), the superposition weight $A$ can be expressed in terms of the initial conditions as

$$A = \frac{1 + \dot{\bar{x}}(0)}{1 - \dot{\bar{x}}(0)}\, e^{-2i\bar{x}(0)}. \tag{18}$$

Writing $\bar{x} = \bar{x}_R + i\bar{x}_I$ and $A = A_r e^{iA_\theta}$, with $\bar{x}_R, \bar{x}_I, A_r, A_\theta \in \mathbb{R}$, Eq. (16) can be rearranged into the implicit constraint

$$\cosh(A_r - 2\bar{x}_I) = \cos(A_\theta + 2\bar{x}_R) + \frac{1}{2}\left|\frac{D^2}{A}\right|. \tag{19}$$

For unbounded propagation, the imaginary coordinate $\bar{x}_I$ must be single-valued for arbitrary $\bar{x}_R$. Since

$\cosh(A_r - 2\bar{x}_I) \geq 1$, a sufficient condition is

$$\frac{1}{2}\left|\frac{D^2}{A}\right| \geq 2. \tag{20}$$

This condition can be rewritten as a constraint on the initial complex velocity,

$$|\dot{\bar{x}}_R(0)|^2 - |\dot{\bar{x}}_I(0)|^2 \geq \frac{1}{2}. \tag{21}$$

Accordingly, the photon motion can be classified into three modes:

$$\bar{x}_+(t): \dot{\bar{x}}_R(0) > \sqrt{\dot{\bar{x}}_I^2(0) + \frac{1}{2}} \Rightarrow \text{right-propagating}, \tag{22a}$$

$$\bar{x}_b(t): |\, \dot{\bar{x}}_R(0) \,| < \sqrt{\dot{\bar{x}}_I^2(0) + \frac{1}{2}} \Rightarrow \text{bounded}, \tag{22b}$$

$$\bar{x}_-(t): \dot{\bar{x}}_R(0) < -\sqrt{\dot{\bar{x}}_I^2(0) + \frac{1}{2}} \Rightarrow \text{left-propagating}. \tag{22c}$$

In the propagating modes, the complex trajectories are oscillatory, while the bounded mode corresponds to a closed loop in the complex plane as Fig. 1 displays. All three trajectories arise from superposition states and are inherently periodic, highlighting the geometric origin of wave-like behavior in complex space.

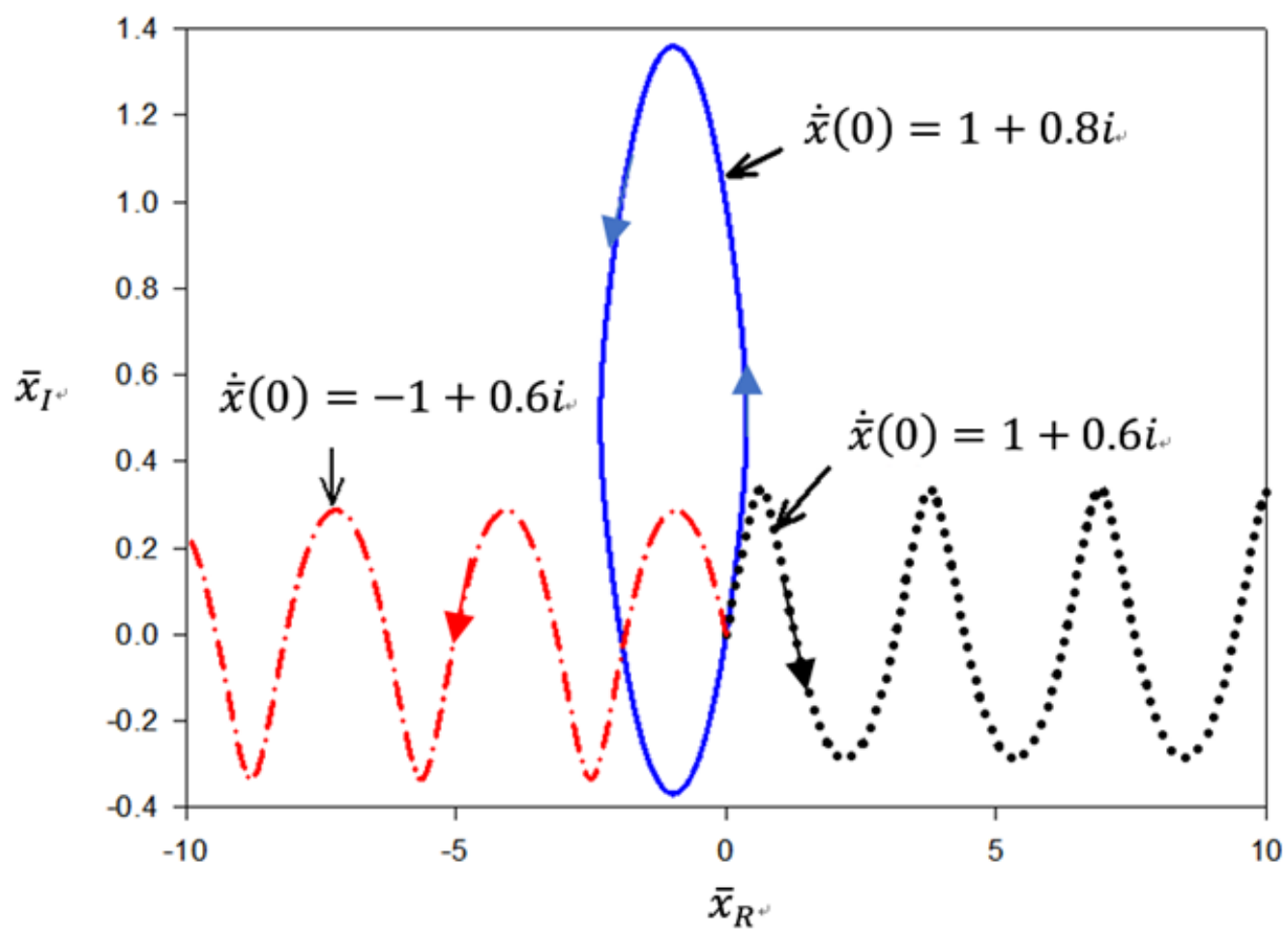

**Figure 1.** Three representative photon trajectories in the complex one-dimensional configuration space $\bar{x} = \bar{x}_R + i\bar{x}_I$, obtained from the relativistic quantum Hamilton–Jacobi equations. All trajectories start from the same initial position but differ in their initial complex velocities. The right- and left-propagating modes correspond to unbounded motion with oscillatory structure in the complex plane, while the bounded mode forms a closed loop. These trajectories arise for superposition states and provide a geometric representation of wave-like behavior in complex space.

For the right-propagating mode, the dimensionless period and wavelength associated with the oscillatory trajectory are given by $\bar{T} = \pi,\ \bar{\lambda} = \pi$ as Figures 2 shows. Upon converting to dimensional variables, one obtains

$$\lambda = \frac{\pi}{k_1}, \qquad T = \frac{\pi}{ck_1}, \tag{23}$$

which leads directly to

$$v = \frac{\lambda}{T} = c. \tag{24}$$

This result demonstrates that the photon propagates at the speed of light regardless of whether it is described by a particle-like eigenstate or a wave-like superposition state. Figures 2 further quantifies the wave-like properties of the right-propagating mode. As shown in Figure 2(a), increasing the imaginary component of the initial complex velocity, $\bar{x}_I(0)$, results in a larger oscillation amplitude of the trajectory. Simultaneously, Figure 2(b) shows that a larger $\bar{x}_I(0)$ produces a greater deviation of the real component $\bar{x}_R(t)$ from a straight-line path. In the limiting case $\bar{x}_I(0) \to 0$, the oscillatory component vanishes and the trajectory reduces to uniform straight-line propagation along the real axis, recovering particle-like motion in a momentum eigenstate. Within this framework, the wave character of the photon does not arise from a classical field oscillation, but rather from oscillatory motion in the imaginary direction of its complex trajectory.

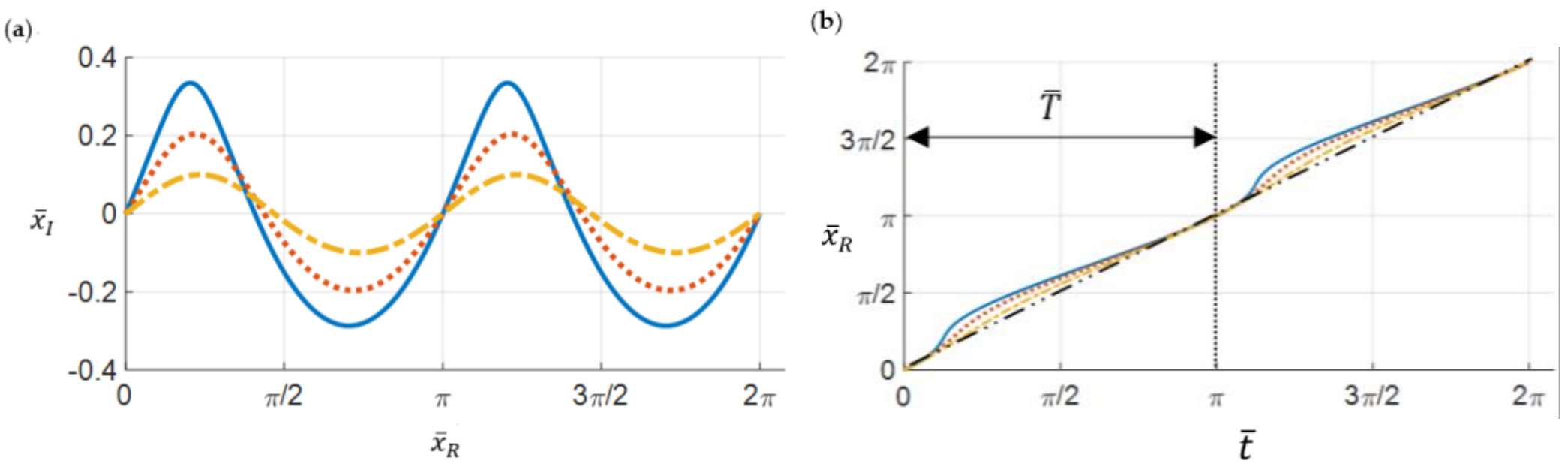

**Figure 2. (a)** Complex photon trajectories in the right-propagating mode for different initial imaginary velocity components $\dot{\bar{x}}_I(0)$. Increasing the imaginary component enhances the oscillation amplitude in the complex plane. **(b)** Corresponding real-space projections $\bar{x}_R(t)$. As $\dot{\bar{x}}_I(0)$ increases, deviations from straight-line propagation become more pronounced, while the average propagation speed remains equal to the speed of light.

From the Hamiltonian

$$H = \frac{c^2}{E_0}(p_0^2 - p_1^2) + \frac{c^2\hbar}{E_0 i}\left(\frac{\partial p_0}{c\,\partial t} - \frac{\partial p_1}{\partial x}\right), \tag{25}$$

the associated complex quantum potential is

$$Q = \frac{c^2\hbar}{E_0 i}\left(\frac{\partial p_0}{\partial(ct)} - \frac{\partial p_1}{\partial x}\right) = \frac{c^2\hbar^2}{E_0}\left(-\frac{\partial^2}{c^2\,\partial t^2}\ln\,\Psi + \frac{\partial^2}{\partial x^2}\ln\,\Psi\right). \tag{26}$$

For a single momentum eigenstate, $Q = 0$ and the trajectory is straight. For a superposition state, the quantum potential is nonzero. In dimensionless form,

$$\bar{Q} = \frac{2A}{(Ae^{i\bar{x}} + e^{-i\bar{x}})^2}, \tag{27}$$

and the corresponding force term is

$$\bar{F} = -\frac{\partial \bar{Q}}{\partial \bar{x}} = \frac{4Ai(Ae^{i\bar{x}} - e^{-i\bar{x}})}{(Ae^{i\bar{x}} + e^{-i\bar{x}})^3}. \tag{28}$$

The bounded trajectories can therefore be interpreted as resulting from effective trapping by the quantum potential in the complex plane, with turning points located near regions of high potential. Figures (3a) and (3b) illustrates how does the photon follow the quantum potential variations to change its motion and how the quantum force affect it. Figure 4 displays how the quantum potential traps the photon in the potential well, allowing the photon to have a bounded trajectory with the initial condition given by $\left(\dot{x}_R(0), \dot{x}_I(0)\right) = (1,0.8)$. The high potential barriers at two edges are the locations where turning points of the complex trajectory occur.

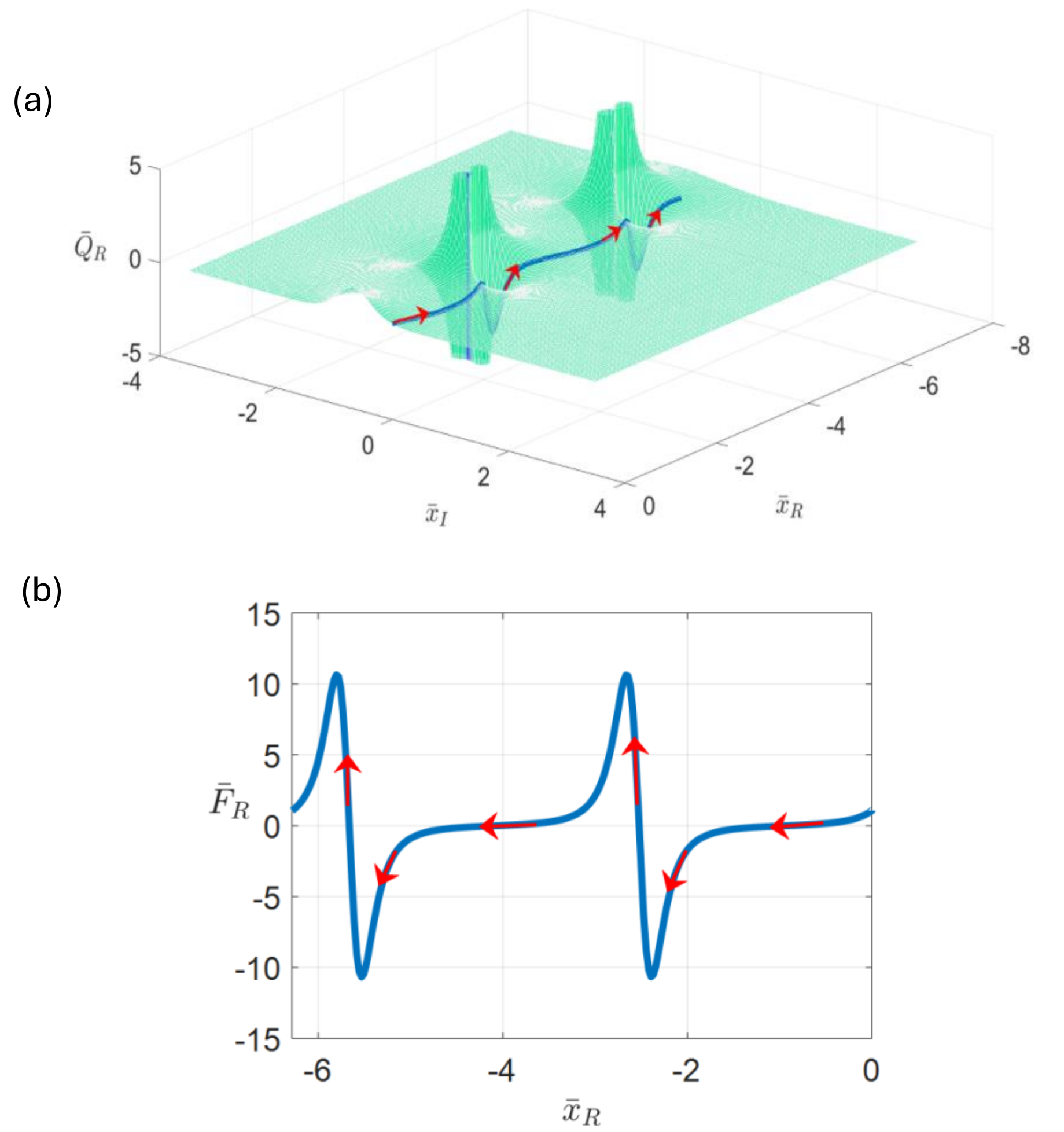

**Figure 3. (a)** A complex photon trajectory superimposed on the real part of the quantum potential along the real axis. **(b)** The corresponding quantum force derived from the gradient of the quantum potential. The turning points of the trajectory coincide with regions of large potential gradients, indicating effective confinement in complex space.

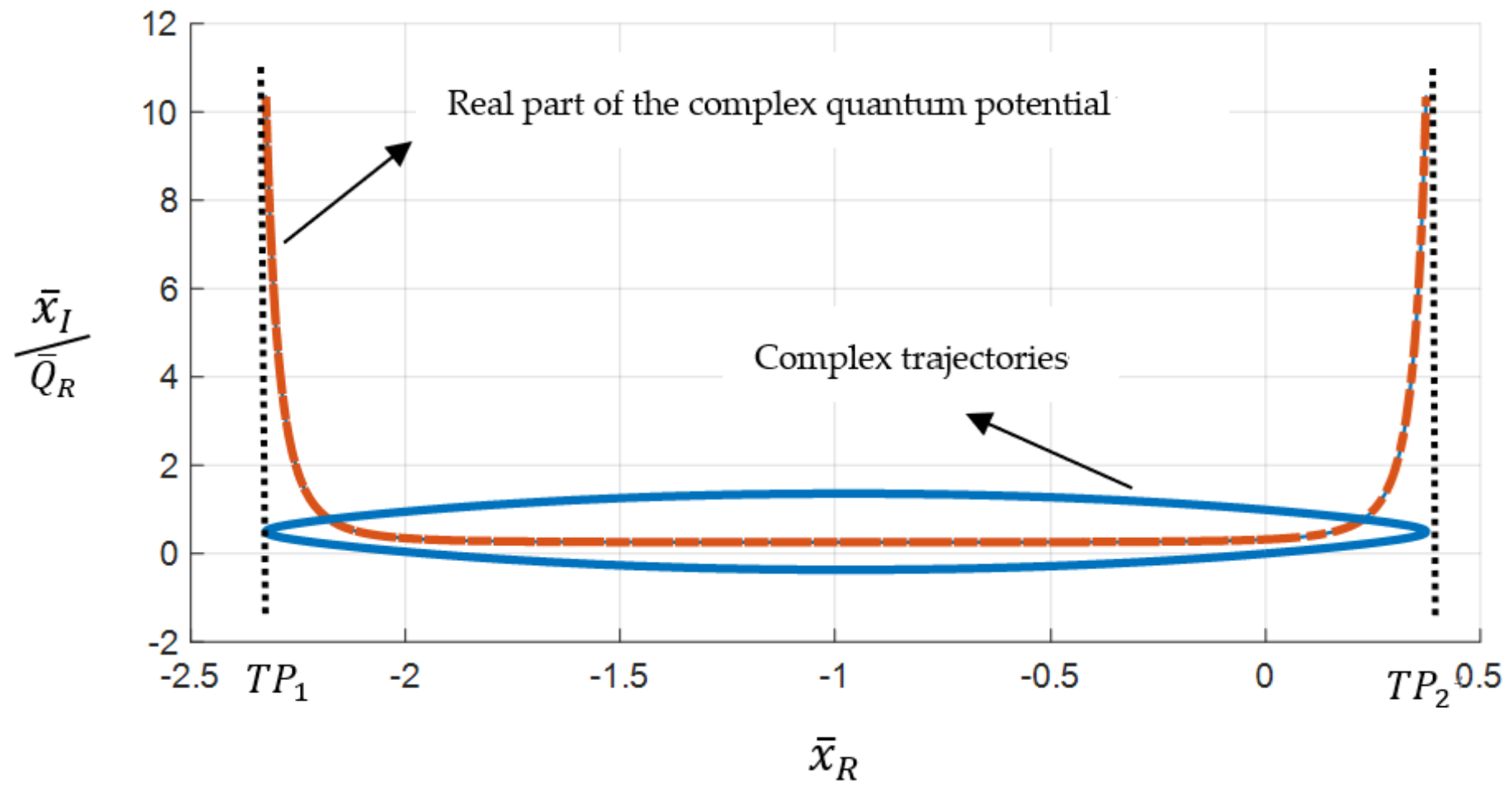

**Figure 4.** Bounded photon motion in the complex plane resulting from an effective potential well formed by the quantum potential. The turning points of the trajectory correspond to high-potential regions, illustrating how bounded dynamics arise naturally for certain superposition states.

## 4. Single Photon Motions in Complex Space

We extend the one-dimensional analysis to complex two-dimensional space, $(\bar{x}, \bar{y}) \in \mathbb{C}^2$, and consider a free photon described by the separable superposition state

$$\Psi(\bar{t}, \bar{x}, \bar{y}) = e^{i\bar{t}}(Ae^{i\bar{x}} + e^{-i\bar{x}})(Be^{i\bar{y}} + e^{-i\bar{y}}), \bar{x}, \bar{y} \in \mathbb{C}. \tag{29}$$

Here $A = C_x^+/C_x^-$ and $B = C_y^+/C_y^-$ are the relative weights of the $+$ and $-$momentum components along the $\bar{x}$ and $\bar{y}$ directions. Introducing the dimensionless variables

$$\bar{t} = \frac{k_0}{\hbar}t, \qquad \bar{x} = k_1 x, \qquad \bar{y} = k_2 y, \tag{30}$$

and defining the normalized directional factors

$$n_x = \frac{\hbar c k_1}{k_0}, \qquad n_y = \frac{\hbar c k_2}{k_0}, \quad n_x^2 + n_y^2 = 1, \tag{31}$$

where the last relation follows from the massless dispersion $k_0^2 = \hbar^2 c^2 (k_1^2 + k_2^2)$, the equations of motion separate into two independent complex equations,

$$\frac{d\bar{x}}{d\bar{t}} = n_x \frac{Ae^{i\bar{x}} - e^{-i\bar{x}}}{Ae^{i\bar{x}} + e^{-i\bar{x}}}, \tag{32a}$$

$$\frac{d\bar{y}}{d\bar{t}} = n_y \frac{Be^{i\bar{y}} - e^{-i\bar{y}}}{Be^{i\bar{y}} + e^{-i\bar{y}}}. \tag{32b}$$

As in the one-dimensional case, the superposition weights $A$ and $B$ can be parameterized by the initial conditions. Writing the initial complex velocities as $\dot{\bar{x}}(0)$and $\dot{\bar{y}}(0)$, one obtains

$$A = \frac{1 + \dot{\bar{x}}(0)/n_x}{1 - \dot{\bar{x}}(0)/n_x}\, e^{-2i\bar{x}(0)}, \tag{33a}$$

$$B = \frac{1 + \dot{\bar{y}}(0)/n_y}{1 - \dot{\bar{y}}(0)/n_y}\, e^{-2i\bar{y}(0)}. \tag{33b}$$

Integrating Eqs. (32) yields implicit solutions of the form

$$Ae^{i\bar{x}} - e^{-i\bar{x}} = D\, e^{-i\bar{t}}, \tag{34a}$$

$$Be^{i\bar{y}} - e^{-i\bar{y}} = J\, e^{-i\bar{t}}, \tag{34b}$$

with integration constants fixed by initial conditions,

$$D = \frac{2\dot{\bar{x}}(0)/n_x}{1 - \dot{\bar{x}}(0)/n_x}\, e^{-i\bar{x}(0)}, \tag{35a}$$

$$J = \frac{2\dot{\bar{y}}(0)/n_y}{1 - \dot{\bar{y}}(0)/n_y}\, e^{-i\bar{y}(0)}. \tag{35b}$$

Each complex plane admits three qualitative modes (right/left propagation and bounded motion), determined by the initial complex velocity. A convenient sufficient criterion for unbounded propagation in the $\bar{x}$ −direction is

$$\dot{\bar{x}}_R^2(0) - \dot{\bar{x}}_I^2(0) > \frac{n_x^2}{2}, \tag{36a}$$

with the complementary condition corresponding to bounded motion,

$$\dot{\bar{x}}_R^2(0) - \dot{\bar{x}}_I^2(0) \le \frac{n_x^2}{2}. \tag{36b}$$

Analogously, for the $\bar{y}$-direction,

$$\dot{\bar{y}}_R^2(0) - \dot{\bar{y}}_I^2(0) > \frac{n_y^2}{2}, \tag{36c}$$

$$\dot{\bar{y}}_R^2(0) - \dot{\bar{y}}_I^2(0) \le \frac{n_y^2}{2}. \tag{36d}$$

Combining the mode choice in $\bar{x}$ and $\bar{y}$ yields four representative two-dimensional behaviors: (a) propagation in both directions; (b) propagation in $\bar{y}$ with bounded motion in $\bar{x}$; (c) propagation in $\bar{x}$ with bounded motion in $\bar{y}$; and (d) bounded motion in both directions.

For the numerical demonstrations reported below, we adopt $n_x = \sqrt{1/8}$, $n_y = \sqrt{7/8}$, and the initial position

$$(\bar{x}_R(0), \bar{x}_I(0), \bar{y}_R(0), \bar{y}_I(0)) = (0,0,0,0), \tag{37}$$

while Table.1 specifies the initial complex velocities used to illustrate the four mode combinations. In the simulations, three characteristic trajectory morphologies appear depending on the initial complex velocities: (i) weakly modulated propagation, (ii) strongly modulated propagation, and (iii) standing-wave-like patterns in the real $\bar{x}_R$–$\bar{y}_R$ projection.

Qualitatively, when propagation occurs in both directions (mode a), the real-plane trajectory is close to a straight line with small oscillatory deviations; the corresponding complex-plane projections $(\bar{x}_R, \bar{x}_I)$ and $(\bar{y}_R, \bar{y}_I)$ exhibit oscillations driven by the nonzero imaginary parts of the complex velocities. When one direction is bounded and the other propagates (modes b and c), the bounded direction exhibits a larger-amplitude oscillation, leading to more pronounced wave-like modulation in the real projection. When both directions are bounded (mode d), the real-plane projection becomes standing-wave-like over a finite region, consistent with confinement by the complex quantum potential discussed below.

Figure 5(a) shows the photon trajectory in the real $\bar{x}$–$\bar{y}$ plane for mode (a), where the photon propagates freely in both directions. The trajectory is predominantly linear, with small deviations from a straight line that manifest as a weakly wave-like structure. These deviations originate from the oscillatory motion in the associated complex planes. As illustrated in Figure 5(b), the trajectories in the complex $\bar{x}$- and $\bar{y}$-planes exhibit clear wave-like oscillations, which arise from the nonzero imaginary components of the initial complex velocity. Notably, the oscillation frequency in the complex $\bar{y}$-plane is higher than that in the complex $\bar{x}$-plane, reflecting a larger energy contribution in the $+\bar{y}$-direction. This anisotropy is consistent with the condition $n_y > n_x$.

When the photon propagates along one direction while remaining bounded in the orthogonal direction, the oscillatory amplitude becomes more pronounced in the bounded coordinate, resulting in an apparent wave-like trajectory in the real plane. Figure 6(a) depicts the photon trajectory in the real $\bar{x}$–$\bar{y}$ plane for mode (b). As shown in Figure 6(b), the photon propagates along the $+\bar{y}$-direction in the complex $\bar{y}$-plane, while its motion in the complex $\bar{x}$-plane is bounded. This behavior arises from the initial complex velocity satisfying conditions (36b) and (36c), respectively.

A complementary configuration is presented in Fig. 7 for mode (c), where the photon propagates along the $+\bar{x}$-direction and is bounded in the $+\bar{y}$-direction. Figure 7(a) displays the resulting apparent wave-like trajectory in the real $\bar{x}$–$\bar{y}$ plane, while Fig. 7(b) confirms the bounded oscillatory motion in the complex $\bar{y}$-plane.

Finally, when the photon motion is bounded in both the $+\bar{x}$- and $+\bar{y}$-directions, corresponding to mode (d), a standing-wave pattern emerges in the real $\bar{x}$–$\bar{y}$ plane. As shown in Fig. 8, the photon starts at $\bar{x}_R = 0$, undergoes oscillatory motion with finite amplitude in the $\bar{y}_R$-direction, and completes one full period by returning to $\bar{x}_R = 0$ at $\bar{t} = 16\pi$. This standing-wave behavior highlights the direct correspondence between bounded complex trajectories and stationary wave patterns in real space.

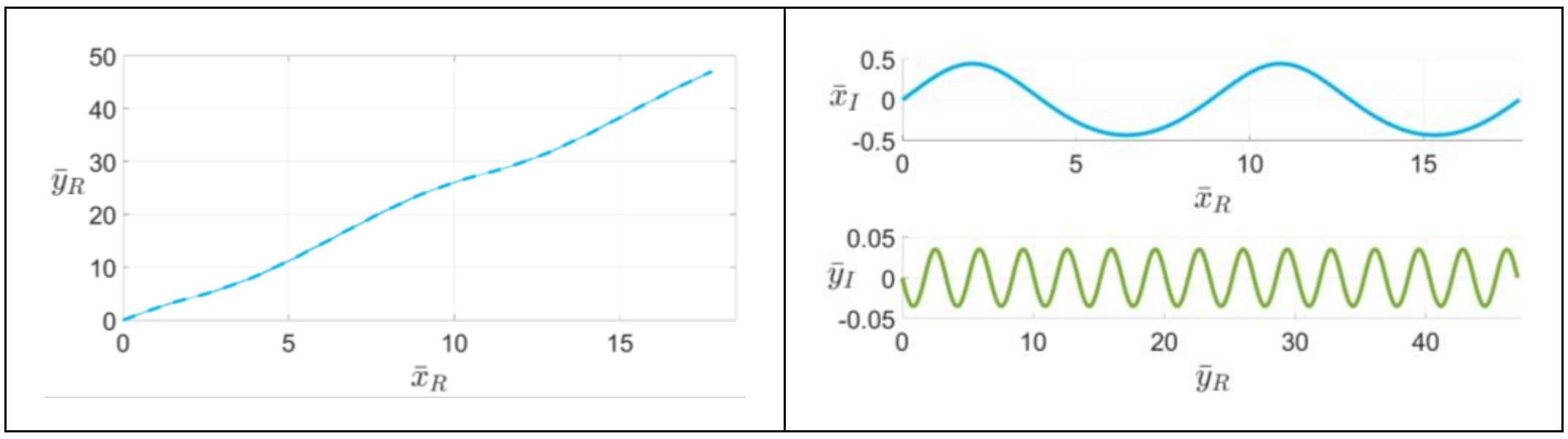

**Figure 5.** The photon is in the propagation mode in both $+\bar{x}-$ and $+\bar{y}-$ directions with the energy distributions $n_x = \sqrt{1/8}$ and $n_y = \sqrt{7/8}$. The initial position is $(\bar{x}_R(0), \bar{x}_I(0), \bar{y}_R(0), \bar{y}_I(0)) = (0,0,0,0)$, and the initial complex velocity is assigned as $\left(\dot{\bar{x}}_R(0), \dot{\bar{x}}_I(0), \dot{\bar{y}}_R(0)\right) = \left(\sqrt{1/8}, 0.11, \sqrt{7/8}, -0.0614\right)$. (a) The real-space projection exhibits slight wave-like modulation superimposed on a nearly straight trajectory. (b) The wave motions that the photon has in the complex $\bar{x}$ −plane and complex $\bar{y}$ −plane.

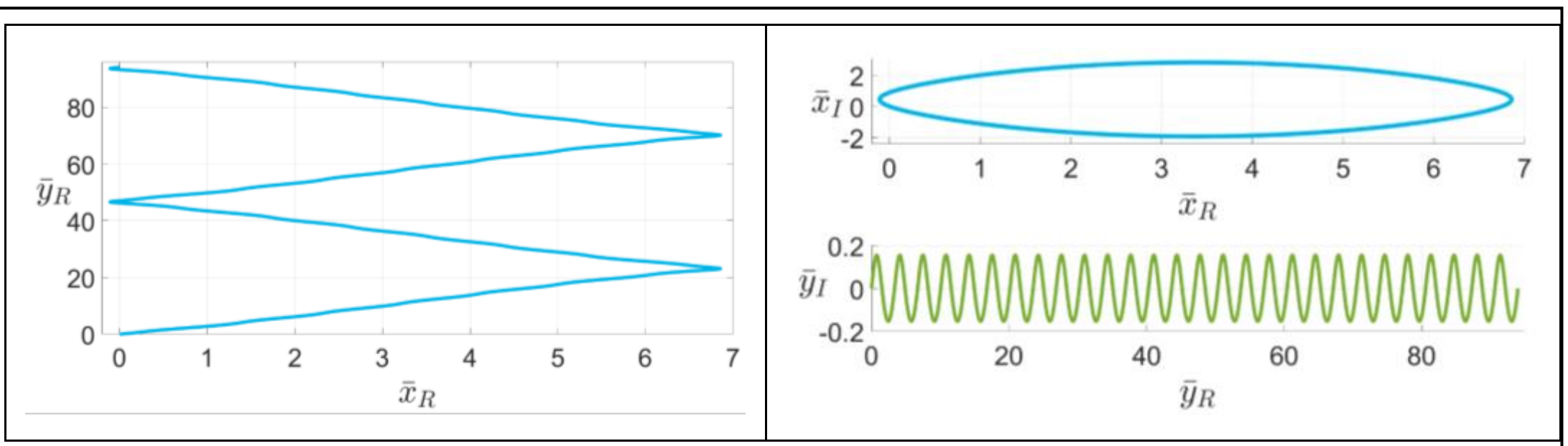

**Figure 6.** Photon trajectories propagating in one direction while remaining bounded in the orthogonal complex direction. The bounded motion gives rise to enhanced oscillatory patterns in the real-space projection. The photon is in the propagation mode in $+\bar{y}-$ direction and in bounded mode in $+\bar{x}-$ direction with the energy distributions $n_x = \sqrt{1/8}$ and $n_y = \sqrt{7/8}$. The initial position is $(\bar{x}_R(0), \bar{x}_I(0), \bar{y}_R(0), \bar{y}_I(0)) = (0,0,0,0)$, and the initial complex velocity is assigned as $\left(\dot{\bar{x}}_R(0), \dot{\bar{x}}_I(0), \dot{\bar{y}}_R(0)\right) = \left(\sqrt{1/8}, -0.75, \sqrt{7/8}, -0.2714\right)$. (a) The photon motion with apparent wave-like in the real $\bar{x} - \bar{y}$ plane. (b) The closed trajectory and wave motions that the photon has in the complex $\bar{x}$ −plane and $\bar{y}$ −plane, respectively.

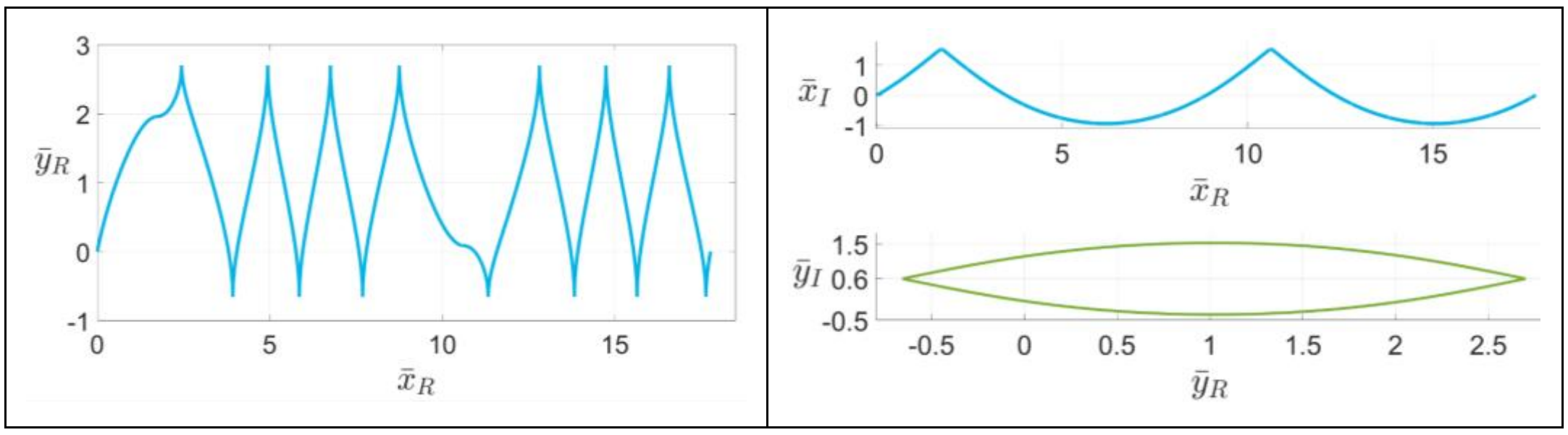

**Figure 7.** Apparent wave-like trajectory. The photon is in the propagation mode in the $+\bar{x}-$ direction and in bounded mode in $+\bar{y}-$ direction with the energy distributions $n_x = \sqrt{1/8}$ and $n_y = \sqrt{7/8}$. The initial position is $(\bar{x}_R(0), \bar{x}_I(0), \bar{y}_R(0), \bar{y}_I(0)) = (0,0,0,0)$, and the initial complex velocity is assigned as $\left(\dot{\bar{x}}_R(0), \dot{\bar{x}}_I(0), \dot{\bar{y}}_R(0)\right) = \left(\sqrt{1/8}, 0.24995, \sqrt{7/8}, -0.6614\right)$. (a) The photon motion with apparent wave-like in the real $\bar{x}-\bar{y}$ plane. (b) The wave motions and closed trajectory that the photon has in the complex $\bar{x}$ −plane and $\bar{y}$ −plane, respectively.

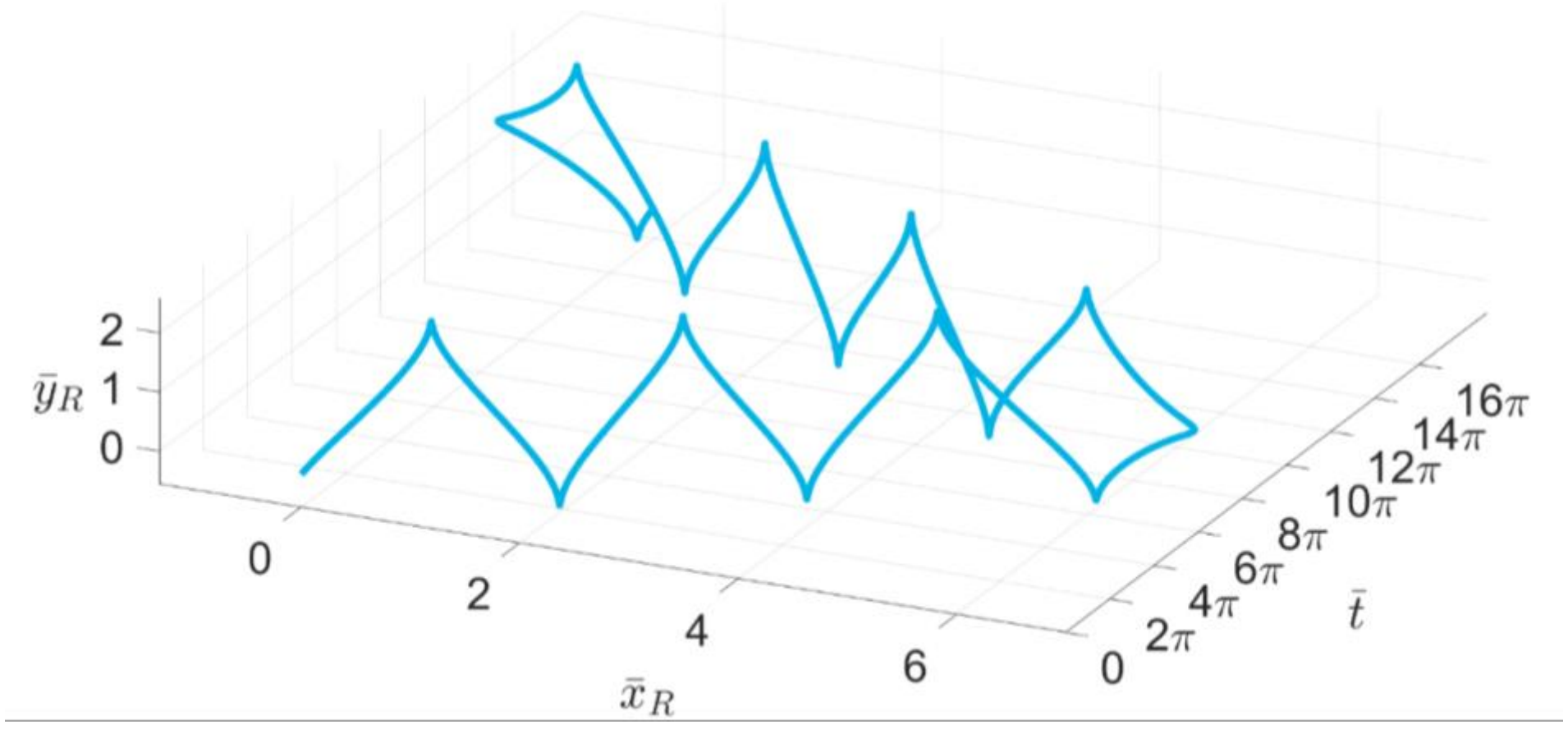

**Figure 8.** Standing-wave-like patterns in the real $\bar{x}_R - \bar{y}_R$ plane resulting from bounded motion in both complex directions. The standing-wave structure arises as a projection of bounded complex trajectories rather than from a static wave field. The photon is in the bounded mode in $+\bar{x}$ and $+\bar{y}-$ directions with the energy distributions $n_x = \sqrt{1/8}$ and $n_y = \sqrt{7/8}$. The initial position is $(\bar{x}_R(0), \bar{x}_I(0), \bar{y}_R(0), \bar{y}_I(0)) = (0,0,0,0)$, and the initial complex velocity is assigned as $\left(\dot{\bar{x}}_R(0), \dot{\bar{x}}_I(0), \dot{\bar{y}}_R(0)\right) = \left(\sqrt{1/8}, -0.25, \sqrt{7/8}, -0.6614\right)$.

The relativistic quantum Hamilton-Jacobi formulation introduces an effective quantum-potential term whose value vanishes for a single eigenstate but becomes nonzero for superpositions. In the two-dimensional setting, the nontrivial potential structure leads naturally to bounded trajectories in one or both directions, which can be interpreted as confinement within an effective potential well in $\mathbb{C}^2$. To test internal consistency of the standing-wave-like regime (mode d), we evaluate effective wavelengths extracted from the projected standing-wave patterns. For example, from the standing-wave-like trajectory in the real $\bar{x}_R$–$\bar{y}_R$ projection, a dimensionless wavelength $\bar{\lambda}_{(x)}$ can be estimated along the $\bar{x}_R$-direction. Converting to dimensional wavelengths gives

$$\lambda_{(x)} = \frac{\bar{\lambda}_{(x)}}{k_1},\ \ \lambda_{(y)} = \frac{\bar{\lambda}_{(y)}}{k_2}. \tag{38}$$

Using the massless relations implied by Eq. (31),

$$k_1 = \frac{n_x}{\hbar c} k_0,\ \ k_2 = \frac{n_y}{\hbar c} k_0, \tag{39}$$

and the photon energy $k_0 = pc = hc/\lambda$, one may compare the inferred energies associated with different projected wavelengths to the source photon energy.

Basically, the photon in the mode (d) is constrained in a region by the complex quantum potential, hence it moves boundedly in four-dimension complex space. Despite it is difficult to demonstrate the photon motion in the four-dimension space, it is easy to be displayed in the projected planes. Figure 10 illustrates all six projected planes of the four-dimension space, includes ①$X_R - X_I$ plane; ② $Y_R - Y_I$ plane; ③ $X_R - Y_R$ plane; ④ $X_I - Y_R$ plane; ⑤ $X_I - Y_I$ plane; ⑥ $X_R - Y_I$ plane. The photon motion in the mode (d) projected onto six planes are shown in Figures 10. The bounded trajectories are observed in the $\bar{x}_R - \bar{x}_I$ plane and $\bar{y}_R - \bar{y}_I$ in Figures 10 (a) and (b); while the standing waves appear in the rest planes. In general, we can regard the bounded trajectory as a kind of the standing wave, then all six projected trajectories are standing waves. It seems that the complex quantum potential forms a potential well in the complex space and keeps the photon moving within it. What causes our attention is that does the energy conserve in the mode (d)? The answer should be yes if the relativistic quantum Hamilton-Jacobi equation is correct. Let us find it out by presenting the energy calculation in the complex space.

The dimensionless wavelength of the standing wave, for example, $\bar{\lambda}_{③} = 2(|\mathrm{Max}(\bar{x}_R)| + |\mathrm{Min}(\bar{x}_R)|) = 15.9968$ is calculated along the $\bar{x}_R -$ direction from the standing wave in the $\bar{x}_R - \bar{y}_R$ plane, as Figure 10(c) illustrates. The dimensional value of $\bar{\lambda}_{③}$ can be obtained via

$$\lambda_{(x)} = \frac{\bar{\lambda}_R}{k_1} = \left(\frac{\sqrt{8}}{2\pi}\lambda\right)\bar{\lambda}_{(x)}, \tag{40a}$$

$$\lambda_{(y)} = \frac{\bar{\lambda}_I}{k_2} = \left(\frac{1}{2\pi}\frac{\sqrt{8}}{\sqrt{7}}\lambda\right)\bar{\lambda}_{(y)}, \tag{40b}$$

where $n_x = \sqrt{1/8}$ and $n_y = \sqrt{7/8}$ are applied to $k_1$ and $k_2$:

$$k_1 = \frac{1}{\sqrt{8}}\frac{k_0}{\hbar c} = \frac{1}{\sqrt{8}}\frac{1}{\hbar c}\frac{hc}{\lambda} = \frac{2\pi}{\sqrt{8}\lambda}, \tag{41a}$$

$$k_2 = \frac{\sqrt{7}}{\sqrt{8}}\frac{k_0}{\hbar c} = \frac{\sqrt{7}}{\sqrt{8}}\frac{1}{\hbar c}\frac{hc}{\lambda} = \frac{\sqrt{7}}{\sqrt{8}}\frac{2\pi}{\lambda}, \tag{41b}$$

in which the total energy of the photon, $k_0 = pc = hc/\lambda$ has been employed. The dimensional relation, Eq. (41b) can be applied to the wavelength calculation along the $\bar{y}_R -$ direction.

The wave energy in each projected plane can be obtained according to its wavelength. If the light source is emitted from the ruby laser with wavelength 694.3 nm, then the wavelengths of the six standing waves can be obtained, as Table 2 shows. The total energy given by the summation of six wave energies,

$$E = E_{x_R x_I} + E_{Y_R Y_I} + E_{x_R y_R} + E_{x_I y_R} + E_{x_I y_I} + E_{x_R y_I}, \tag{42}$$

is 1.7856795 e.v., which is very closed to the source light energy, 1.78568342 e.v. The theoretical energy calculating from the summation of six wave energies in different projected planes has only 0.00026 % percentage error compared to the source light energy. There are five different light sources and energy comparisons listed in Table 3, all point out the energy is conserved in the considered complex space.

There are two formulas we used in our calculation: Einstein's photon energy and de Broglie relation:

$$E = hv, \tag{43}$$

$$\lambda = \frac{h}{p}, \tag{44}$$

represent the particle-like property and wave-like property of light. Combining with the Einstein's energy momentum relation, $E = pc$ we can have the following expression:

$$E = hv = pc \Rightarrow p = \frac{h}{\lambda}, \tag{45}$$

by which one can see that the photon's energy given by the light wave source transfers to the standing wave energy in our calculation. Eq. (45) emerges the particle-like property and wave-like property of light from the perspective of the energy. For a century, Einstein's photon energy formula is only for light with particle-like property, and de Broglie relation considers the wave-property only. Through the motion of photons in complex space, we have demonstrated that light is both particle and wave in nature. Furthermore, we have combined the two equations (43) and (44) into one form, in which waves have particle properties and particles have wave properties.

Within the present framework, the standing-wave-like projections provide a geometric representation of how particle and wave descriptions can be consistently related through energy–momentum constraints, while the underlying dynamics remain encoded in the complex trajectory structure.

**Table 1.** Initial complex velocities and corresponding trajectory types.

| Trajectory Type | Mode | Motion Modes | Initial Complex Velocity |
|---|---|---|---|
| Slight wave-like | (a) | $+\bar{x}$-propagation<br>$+\bar{y}$-propagation | $\dot{x}_R(0) = n_x,\ \ \dot{x}_I(0) = 0.11$<br>$\dot{y}_R(0) = n_y,\ \ \dot{y}_I(0) = -0.0614$ |
| Apparent wave-like | (b) | $+\bar{x}$-bounded<br>$+\bar{y}$-propagation | $\dot{x}_R(0) = n_x,\ \ \dot{x}_I(0) = -0.75$<br>$\dot{y}_R(0) = n_y,\ \ \dot{y}_I(0) = -0.2714$ |
| | (c) | $+\bar{x}$-propagation | $\dot{x}_R(0) = n_x,\ \ \dot{x}_I(0) = 0.24995$ |
| | | $+\bar{y}$-bounded | $\dot{\bar{y}}_R(0) = n_y,\quad \dot{\bar{y}}_I(0) = -0.6614$ |
| Standing wave | (d) | $+\bar{x}$-bounded<br>$+\bar{y}$-bounded | $\dot{\bar{x}}_R(0) = n_x,\quad \dot{\bar{x}}_I(0) = -0.24699$<br>$\dot{\bar{y}}_R(0) = n_y,\quad \dot{\bar{y}}_I(0) = -0.67642$ |

* $n_x = \sqrt{1/8}$ and $n_y = \sqrt{7/8}$

**Table 2.** Wavelengths and energies extracted from different projection planes. (wavelength is 694.3 nm, energy is 1.78568342 e.v.)

| No. | Wave Type | Plane | Wavelength | Wavelength (nm) | Energy (e.v.) |
|---|---|---|---|---|---|
| 1 | Bounded | $\bar{x}_R - \bar{x}_I$ | 15.9968 | 4999.7 | 0.24797434 |
| 2 | Bounded | $\bar{y}_R - \bar{y}_I$ | 44.06 | 5204.8 | 0.23820099 |
| 3 | Standing | $\bar{x}_R - \bar{y}_R$ | 15.9968 | 4999.7 | 0.24797434 |
| 4 | Standing | $\bar{x}_I - \bar{y}_R$ | 9.8731 | 3085.8 | 0.40177734 |
| 5 | Standing | $\bar{x}_I - \bar{y}_I$ | 9.8731 | 3085.8 | 0.40177734 |
| 6 | Standing | $\bar{x}_R - \bar{y}_I$ | 15.9968 | 4999.7 | 0.24797434 |

**Table 3.** Comparison between reconstructed total energy and source photon energy.

| Light Source | Wavelength (nm) | Energy (e.v.) | Theoretical Energy (e.v.) | Percentage Error (%) |
|---|---|---|---|---|
| Ruby laser | 694.3 | 1.78568342 | 1.7856795 | 0.00026 |
| IR diode laser | 850 | 1.45858823 | 1.45858438 | 0.00026 |
| Fiber diode laser | 975 | 1.27158974 | 1.27158638 | 0.00026 |
| Yb:YAG laser | 1030 | 1.20368932 | 1.20368614 | 0.00026 |
| Nb:YVO4 | 1064 | 1.16522556 | 1.16522248 | 0.00026 |

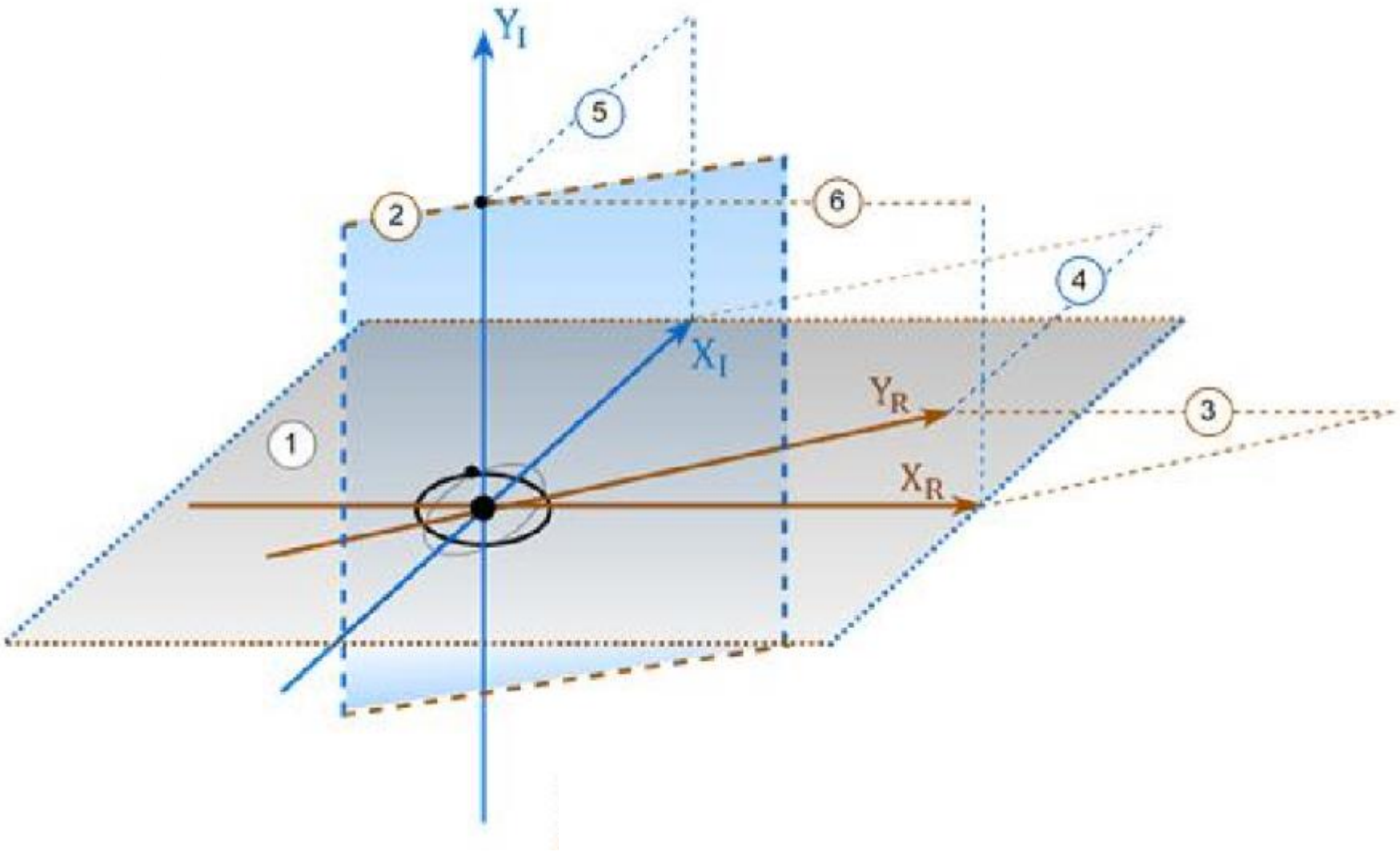

**Figure 9.** Six projection planes of the complex two-dimensional configuration space. The bounded motion in complex space gives rise to standing-wave-like structures in multiple real and mixed projections. ① XR-XI plane; ② YR-YI plane; ③ XR-YR plane; ④ XI-YR plane; ⑤ XI-YI plane; ⑥ XR-YI plane. The axes expressed in brown and blue color represent the real and imaginary axes, respectively.

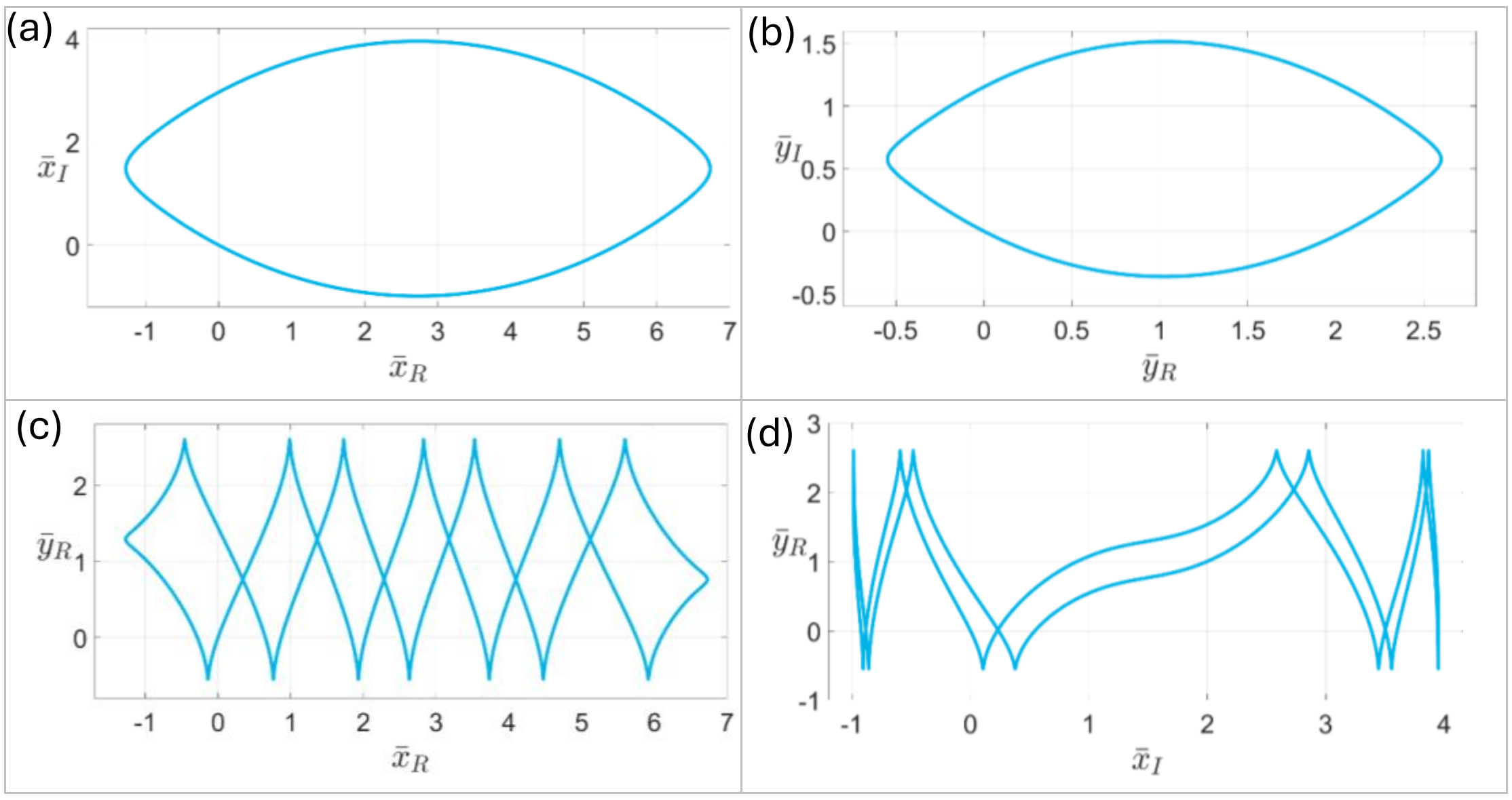

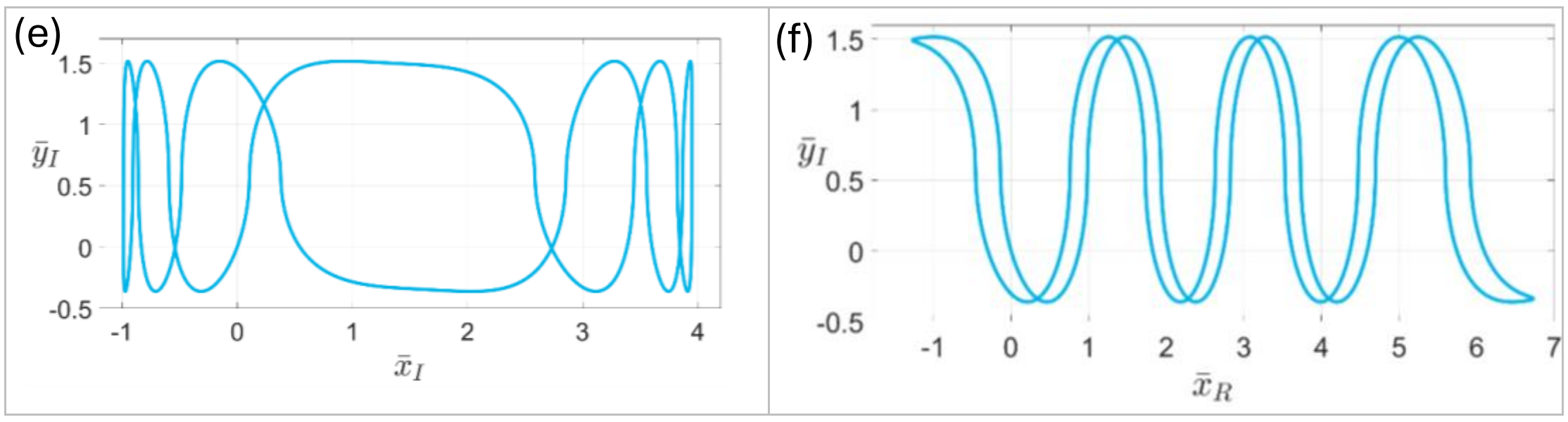

**Figure 10.** Standing-wave–like photon trajectories projected onto different coordinate planes in the complex space. Effective wavelengths are extracted from each projection and used to examine the consistency between the trajectory-derived wave properties and the source photon energy. **(a)** Projection of the photon trajectory onto the $\bar{x}_R$–$\bar{x}_I$ plane, denoted as wave ①, with an effective wavelength $\lambda_1 = 15.9968$ determined from the oscillation along $\bar{x}_R$. **(b)** Projection onto the $\bar{y}_R$–$\bar{y}_I$ plane, denoted as wave ②, with an effective wavelength $\lambda_2 = 44.06$ determined from $\bar{y}_R$. **(c)** Projection onto the real $\bar{x}_R$–$\bar{y}_R$ plane, denoted as wave ③, yielding an effective wavelength $\lambda_3 = 15.9968$ extracted from $\bar{x}_R$. **(d)** Projection onto the $\bar{x}_I$–$\bar{y}_R$ plane, denoted as wave ④, with an effective wavelength $\lambda_4 = 9.8731$ determined from the oscillation along $\bar{x}_I$. **(e)** Projection onto the $\bar{x}_I$–$\bar{y}_I$ plane, denoted as wave ⑤, yielding an effective wavelength $\lambda_5 = 15.9968$ extracted from $\bar{x}_I$. **(f)** Projection onto the $\bar{x}_R$–$\bar{y}_I$ plane, denoted as wave ⑥, with an effective wavelength $\lambda_6 = 15.9968$ determined from $\bar{x}_R$.

## 5. Conclusions

Wave–particle duality continues to motivate efforts to clarify how particle-like and wave-like descriptions arise within a single dynamical framework. In this article we have explored a trajectory-based description formulated in complex space and applied it to the dynamics of a free photon. Within the proposed relativistic quantum Hamiltonian formulation, the photon trajectory is represented as a complex path whose real projection captures propagation while the imaginary components encode oscillatory structure.

In the one-dimensional analysis, a momentum eigenstate (pure right- or left-moving component) yields a straight trajectory and corresponds to uniform propagation at the speed of light. In contrast, superposition states lead to complex velocities and non-straight trajectories in the complex plane. The resulting oscillatory motion in the imaginary direction provides a geometric representation of wave-like behaviour, while the real projection continues to describe propagation along the physical coordinate. In this sense, the distinction between particle-like and wave-like manifestations is traced to whether the effective quantum potential vanishes (eigenstates) or becomes nontrivial (superpositions).

Extending the analysis to complex two-dimensional space further reveals a richer set of behaviours, including propagating wave-like trajectories and standing-wave-like patterns in real projections. When the motion is bounded in both directions, the projected dynamics display standing-wave structure across multiple coordinate planes. As an internal consistency check, we evaluated the energy associated with projected standing-wave wavelengths and found agreement with the source photon energy within numerical accuracy. This supports the view that energy–momentum constraints are maintained within the complex-trajectory description, even when the projected motion appears wave-like.

Finally, the framework provides a unified perspective on standard relations commonly associated with complementary aspects of light. Einstein's photon energy $E = h\nu$ and the de Broglie relation $\lambda = h/p$ appear here as mutually compatible expressions tied together by the relativistic constraint $E = pc$, while the trajectory geometry in complex space determines whether the observable projection emphasizes particle-like transport or wave-like modulation. Rather than treating wave–particle duality as a purely interpretational tension, the complex-trajectory picture suggests a concrete mechanism: superposition induces a nontrivial quantum potential and hence oscillatory structure in the extended complex space.

Several questions remain open. In particular, it will be important to clarify how the present approach interfaces with operational measurement schemes (e.g., weak values), and to test whether the complex-trajectory description can be systematically extended to more general field configurations and multi-mode photonic states. These directions will be pursued in future work. It should be emphasized that the present work is intended as a conceptual and structural investigation rather than a proposal of new dynamical laws or testable predictions. The trajectory-based description developed in this work does not aim to resolve the quantum measurement problem or to provide a complete account of state reduction. The analysis focuses exclusively on free-photon dynamics and idealized superposition states; interactions with measurement apparatuses, environments, or dissipative processes are beyond the present scope.

Finally, the energy-consistency analysis based on projected standing-wave patterns is intended as an internal coherence check rather than an independent derivation of photon energetics. Although the numerical agreement supports the self-consistency of the framework, it does not constitute an experimental validation. Establishing a more direct operational connection—potentially through weak-measurement protocols or related observables—remains an important direction for future investigation.